\documentclass[twocolumn,showpacs,preprintnumbers,amsmath,amssymb,prl,longbibliography,superscriptaddress]{revtex4-1}
\usepackage[scr=boondoxo,scrscaled=1.05]{mathalfa}
\usepackage{tikz}
\usetikzlibrary{calc,decorations.markings}
\usepackage{epsfig}
\usepackage{graphicx}
\usepackage{rotating}
\usepackage{amssymb}
\usepackage{amsmath}
\usepackage{subcaption}
\usepackage{float}   %needed for option H (non floating figure)
\usepackage[font=small,labelfont=bf,justification=justified,format=plain]{caption}
\usepackage[active]{srcltx}
\usepackage{url}
\usepackage{hyperref}
\usepackage{centernot}
\usepackage{dsfont,bm}
\usepackage{xcolor}
\usepackage{pgfplots}
\usepackage{stackengine}
\usepackage{float}  % no floating pictures
\usepackage[scr=boondoxo,scrscaled=1.05]{mathalfa}
\hypersetup{
	colorlinks   = true, %Colours links instead of ugly boxes
	urlcolor     = blue, %Colour for external hyperlinks
	linkcolor    = blue, %Colour of internal links
	citecolor   = red %Colour of citations
}
\usepackage{tikz}
\usetikzlibrary{calc,decorations.markings}
\usepackage{times,graphicx,xcolor}
\usepackage{delarray,fancybox}
\usepackage{mathdots}
\usepackage{eurosym}
\usepackage{relsize}

\usepackage{soul}

\colorlet{darkred}{red!55!black}
\colorlet{darkgreen}{green!25!black}
\newcommand{\ti}{t_{\mathfrak{0}}}

\begin{document}

\title{Interference of Quantum Trajectories}

\author{Brecht Donvil}
\email{brecht.donvil@uni-ulm.de}
\affiliation{Institute for Complex Quantum Systems and IQST, Ulm University - Albert-Einstein-Allee 11, D-89069 Ulm, Germany}
\affiliation{University of Helsinki, Department of Mathematics and Statistics
    P.O. Box 68 FIN-00014, Helsinki, Finland}

\author{Paolo Muratore-Ginanneschi}
\email{paolo.muratore-ginanneschi@helsinki.fi}
\affiliation{University of Helsinki, Department of Mathematics and Statistics
    P.O. Box 68 FIN-00014, Helsinki, Finland}

\begin{abstract}
	We extend quantum trajectory theory to encompass the evolution of a large class of open quantum systems interacting with an environment at {arbitrary coupling strength}. Specifically, we prove that general time-local quantum master equations admit an unraveling described by ordinary jump-stochastic differential equations. The sufficient condition is to weigh the state vector Monte Carlo averages by a probability pseudo-measure which we call the "influence martingale". The influence martingale satisfies a $ 1d $ stochastic differential equation enslaved to the ones governing the quantum trajectories. 
	Our interpretation is that the influence martingale models interference effects between distinct realizations of  quantum trajectories  at strong system-environment coupling.  
	{Our result proves the existence of a Markovian quantum trajectory picture in the Hilbert space of the system for completely bounded divisible dynamical maps. Furthermore, our result provides a new avenue to numerically integrate systems with large numbers of degrees of freedom by naturally extending the existing theory. }
\end{abstract}

\pacs{03.65.Yz, 42.50.Lc}

%\keywords{}
\maketitle

\section{Introduction} 
Actual quantum systems are open: they unavoidably interact, even if slightly, with their surrounding environment \cite{HaRa2006}. 
A useful phenomenological approach is to conceptualize the interaction as a generalized measurement performed by the environment onto the system \cite{BrPe2002, WiMi2009}. As a consequence, the state vector of an open system follows stochastic trajectories in the Hilbert space of its stylized isolated counterpart. 
These trajectories are characterized by sudden transitions, quantum jumps. Since the experimental breakthroughs \cite{SaNeBlTo1986,BeHuItWi1986} quantum jumps have been observed in many atomic and solid-state single quantum systems see e.g. \cite{HaRa2006} for an overview. 
{Quantum trajectory theory \cite{HuPa1984,BaBe1991,GaPaZo1992,DaCaMo1992,Car1993,Bel1994,BrPe1995} (see also \cite{BrPe2002,WiMi2009} for textbook presentations) connects these experimental results to the axiomatic theory of continuous measurement \cite{DaLe1970, Dav1976} which derives deterministic master equations governing the dynamics of a quantum system open also in consequence of the perturbation due to measurement. 
According to quantum trajectory theory, the state operator of the system can be ``unraveled'' i.e. represented as a statistical average over random realizations of post-measurement states.	Mathematically, the average is computed over the realizations of a stochastic process describing the effect on the system of the interaction with an environment subject to continuous monitoring by a measurement device \cite{BaBe1991,Car1993,BrPe2002,Bru2002, JaSt2006,WiMi2009}.	

The precise definition of the stochastic process is contextual to the measurement scheme. Different measurement schemes result in distinct unravelings. In all cases, the stochastic process' evolution law subsumes unitary dynamics with random collapse of the state vector occurring in consequence of an indirect measurement whilst continuously preserving the system's Bloch hyper-sphere. 
Finally, in order to permit to permit a measurement interpretation, the stochastic process must be non-anticipating: the statistics up to the present observation must be invariant with respect to future measurement events \cite{GaWi2002,WiGa2008}}. 

The theory of quantum trajectories is still under active development, see  e.g. \cite{GaMo2014,LuStPi2020}. Recent experiments even support the possibility to use the theory to identify precursors of the imminent occurrence of a jump \cite{MiMuShRe2019}.   

{In the present paper we introduce a general quantum trajectory picture for possibly the most general class of physically relevant time local master equations: the class characterized by fundamental solutions mapping bounded state operators into bounded state operators, i.e. they generate completely bounded maps. We refer the reader to \cite{Pec1994,Pec1995,JoShSu2004,ShSu2005,DoLi2016} for a detailed explanation of the physical relevance of completely bounded maps directly from the postulates of quantum mechanics and to \cite{HaSt2020} for a more phenomenological discussion.}

{Given a microscopic unitary dynamics, a partial trace implemented with the help of time convolutionless perturbation theory \cite{HaShSh1977} (see also e.g. chapters~9 and 10 of \cite{BrPe2002}) yields}
\begin{align}
	\bm{\dot{\rho}}_{t}=-\imath\,\left[\operatorname{H}_t\,,\bm{\rho}_{t}\right] 
	+\sum_{\ell=1}^{\mathscr{L}} \Gamma_{\ell,t}\frac{\left[\operatorname{L}_{\ell}\,,\bm{\rho}_{t}\operatorname{L}_{\ell}^{\dagger}\right]
		+\left[\operatorname{L}_{\ell}\bm{\rho}_{t}\,,\operatorname{L}_{\ell}^{\dagger}\right]}{2}
	\label{LGKS}
\end{align}
{The master equation  (\ref{LGKS}) embodies the universal form of a trace preserving time local evolution law.}
Namely, (\ref{LGKS}) is also obtained by tracing environment degrees of freedom in Gaussian system-environment models of bosons e.g. \cite{FeVe1963,KaGr1997} or Fermions e.g. \cite{TuZh2008,DoGoMG2020}, and in other exactly integrable models of system-environment interactions e.g. \cite{JoQu1994}.

In (\ref{LGKS}), the Hamiltonian $\operatorname{H}_t$ is the generator of a unitary dynamics. The collection $\{\operatorname{L}_{\ell}\}_{\ell= 1}^\mathscr{L}$ consists of so-called Lindblad operators modeling the interaction with the environment. 
The weights  $\Gamma_{\ell,t}$'s are related to the probability per unit of time of the collapse associated to the Lindblad operators they couple to (\ref{LGKS}). Khalfin's theorem \cite{Kha1957} forbids exponential decay in quantum mechanics outside the intermediate asymptotic singled out by the weak coupling scaling limit \cite{Dav1976,FoGhRi1978}. Thus the $ \Gamma_{\ell,t} $'s are in general time dependent functions with arbitrary sign. We only require them to be bounded.  
The celebrated  Lindblad-Gorini-Kossakowski-Sudarshan master equation is thus a special case of (\ref{LGKS}) corresponding to the complete positivity conditions
\begin{align}
	\Gamma_{\ell,t}\,\geq\,0\hspace{1.0cm}\ell=1,\dots,\mathscr{L}.
	\label{CP}
\end{align}
Completely positive dynamics is usually proved from microscopic models only in the weak coupling limit \cite{Dav1976}, see also e.g. \cite{BrPe2002,WiMi2009}. 

{Formulations in the literature of quantum trajectory theory admitting a measurement interpretation also rely on the completely positive conditions (\ref{CP}). Unravelings of the general time local master equation (\ref{LGKS}) require either the introduction of ancillary Hilbert spaces \cite{Bre2004} or to postulate memory and prescience effects between trajectories \cite{PiMaHaSu2008,DiGiSt1998} without a measurement interpretation \cite{Dio2008b,SmCaPi2020}.}

Our main result is to prove that any trace preserving master equation  (\ref{LGKS}) admits an unraveling in non-anticipating quantum trajectories on the system's Bloch hyper-sphere and independently of the sign of  the weights $ \Gamma_{\ell,t} $'s. 
Specifically, we show that the quantum trajectories are realizations of the solutions of a system of ordinary stochastic differential equations driven by independent Poisson processes. 
The only proviso is that each realization of a quantum trajectory enters the Monte-Carlo average with its own weight factor. The weight factor is a martingale $ \mu_{t} $, 
a stochastic process whose expectation value is conserved on average (see e.g. \cite{BoVaWo1975,Kle2005}) so to ensure trace preservation. At any time $ t $, the martingale $ \mu_{t} $  obeys on its turn an ordinary stochastic differential equation enslaved to that governing the state vector of the system.  For reasons that will become clear, we refer to $\mu_{t}$ as the "{influence martingale}". 

{We illustrate our main result in integrable models whose unravelings in the Hilbert space of the system was previously believed to hinge upon memory and prescience effects or simply not possible because of non-positive preserving dynamics (Redfield equation). Besides a measurement interpretation,  unravelings provide a numerical avenue to integrate open quantum systems in high dimensional Hilbert spaces \cite{DaCaMo1992}. In particular, it is well known (see e.g. chapter~7 of \cite{BrPe2002}) that the integration times of the state operator respectively computed from the master equation and from an average over quantum trajectories is expected to undergo a cross-over as the dimension of the Hilbert space increases. We verify 
the existence of the cross-over in a test case using QuTiP \cite{JoNaNo2013}, a widely applied toolbox for efficient numerical simulations of open quantum systems.
We  emphasize that the convergence of unravelings based on the influence martingale is guaranteed by the well established theory of ordinary stochastic differential equations with jumps see e.g. \cite{PlBrLi2010}. }

{Finally we show how the influence martingale naturally accounts for photo-current oscillations which are observed in experimental quantum optics.}

\section{Results}

The gist of the proof of the unraveling of (\ref{LGKS}) via the influence martingale is an elementary extension of Girsanov's change of measure formula, a well known result in the theory of stochastic processes (see e.g. \cite{Kle2005}). 
Roughly speaking, Girsanov formula expresses the average of a generic functional $ F_t $ up to time $ t $ of  a stochastic process in terms of the weighted average of the same  functional now evaluated over a second distinct stochastic process 
\begin{align}
\tilde{\operatorname{E}}	({F}_{t})=\operatorname{E}( {M}_{t}	{F}_{t}).
	\nonumber
\end{align}
Here $ \operatorname{E} $ and $ \tilde{\operatorname{E}} $ denote the expectation values with respect to the probability measures of the two stochastic processes. Girsanov's theorem states that the scalar weighing factor $  {M}_{t}$ must be a positive definite martingale \cite{BoVaWo1975,Kle2005} satisfying for all $ t $ 
\begin{align}
\operatorname{E}( {M}_{t})=1.
	\nonumber
\end{align}
The extension we propose consists in relinquishing the requirement that the martingale be positive definite. 
We will return below on the mathematical and physical interpretation of  the negative values of the influence martingale.  
We now turn to detail the proof of the unraveling. 

{Our aim is to prove} that a state operator solution of  \eqref{LGKS} always admits the representation 
\begin{align}
		\bm{\rho}_{t}=\operatorname{E}\left(\mu_{t}\bm{\psi}_{t}\bm{\psi}_{t}^{\dagger}\right).
	\label{interference}
\end{align}
Here $ \operatorname{E} $ denotes the expectation value operation, $ \bm{\psi}_{t} $ is a stochastic state vector at time $ t $ defined in the Hilbert space of the system and  $\bm{\psi}_{t}^{\dagger}$ is its adjoint dual. 
Finally, $ \mu_{t} $ is a scalar stochastic process enjoying the martingale property. We  prescribe $\bm{\psi}_t$,  $\bm{\psi}_t^{\dagger}$ and $\mu_t$ to obey evolution laws such that the expectation value \eqref{interference} indeed solves  \eqref{LGKS}.

First, we require that the state vector solve the It\^{o} stochastic differential equation \cite{DaCaMo1992}
{
	\begin{align}
	\label{sse}
	\mathrm{d}\bm{\psi}_{t}=&-\imath\,\operatorname{H}_t\bm{\psi}_{t}\mathrm{d}t-
	\sum_{\ell=1}^{\mathscr{L}}\Gamma_{\ell,t}\frac{\operatorname{L}_{\ell}^{\dagger}\operatorname{L}_{\ell}-\left\|\operatorname{L}_{\ell}\bm{\psi}_{t}\right\|^{2}}{2}\bm{\psi}_{t}\mathrm{d}t\nonumber
		\\&+\sum_{\ell=1}^{\mathscr{L}}\mathrm{d}{\nu}_{\ell,t}\left(\frac{\operatorname{L}_{\ell}\bm{\psi}_{t}}{\left\|\operatorname{L}_{\ell}\bm{\psi}_{t}\right\|}-\bm{\psi}_{t}\right)
	\end{align}
}In (\ref{sse}), the $ \left\{  {\nu}_{\ell,t}\right\}_{\ell=1}^{\mathscr{L}} $ are a collection of independent real non-homogeneous Poisson processes (see e.g. \cite{BrPe2002,WiMi2009}). The statistics of the Poisson process increments $\mathrm{d}\nu_{\ell,t}$'s are fully specified for $\ell,\mathscr{k}=1,\dots,\mathscr{L}$ by
{
\begin{subequations}
	\label{Poisson}
	\begin{align}
		&\label{Poisson1}
		\mathrm{d}\nu_{\ell,t}\mathrm{d}\nu_{\mathscr{k},t}=\delta_{\ell,\mathscr{k}}\mathrm{d}\nu_{\ell,t}%\quad \textrm{for}\quad  \ell,\mathscr{k}=1,\dots,\mathscr{L}
		\\
		&\label{Poisson2}
		\operatorname{E}\big{(}\mathrm{d}\nu_{\ell,t}\big{|}\bm{\psi}_{t},\bm{\psi}_{t}^{\dagger}\big{)}
		=\mathscr{r}_{\ell,t}\left\|\operatorname{L}_{\ell}\bm{\psi}_{t}\right\|^{2}\mathrm{d}t
		%&\hspace{-0.5cm}	\ell=1,\dots,\mathscr{L},
	\end{align}
\end{subequations}
}{where $ \left\{ \mathscr{r}_{\ell,t} \right\}_{\ell=1}^{\mathscr{L}} $ is a collection of strictly positive 
definite functions of time.} 
Equation (\ref{Poisson1}) states that $  \mathrm{d}\nu_{\ell,t}$ can only take values $ 0 $ or $ 1 $. The conditional expectation $ \operatorname{E}(\mathrm{d}\nu_{\ell,t}|\bm{\psi}_{t},\bm{\psi}_{t}^{\dagger}) $  determines the jump rates given the values of the state vector and its complex adjoint at time $ t $.
The equations governing $  \bm{\psi}_{t}^{\dagger}$ follow immediately from (\ref{sse}) 
by applying the complex adjoint operation. We associate to (\ref{sse}) and to the equation for the adjoint,  
initial data on the Bloch hyper-sphere i.e. $\bm{\psi}^\dagger_0\bm{\psi}_0=\|\bm{\psi}_0\|^{2}=1$. 
We emphasize that the stochastic Schr\"odinger equation (\ref{sse}) and the noise processes (\ref{Poisson}) are essentially the same as in \cite{DaCaMo1992}.

Next, we need $ \mu_{t} $ to obey an evolution law admitting solutions enjoying the martingale property. We thus require $ \mu_{t}$ to evolve according to the It\^{o} stochastic differential equation
{	
	\begin{subequations}
	\label{martingale}
	\begin{align}
			&\label{martingale1}	\mathrm{d}\mu_{t}=\mu_{t}\sum_{\ell=1}^{\mathscr{L}}\left(\frac{\Gamma_{\ell,t}}{\mathscr{r}_{\ell,t}}-1\right)\left (\mathrm{d}\nu_{\ell,t}-\mathscr{r}_{\ell,t} \left\|\operatorname{L}_{\ell}\bm{\psi}_{t}\right\|^{2}\mathrm{d}t\right )
		\\
		&\label{martingale2}\mu_{0}=1.
	\end{align}
\end{subequations}}
The solution of an equation of the form (\ref{martingale}) is by construction a local martingale (see e.g.  \cite{BoVaWo1975,Kle2005} for details). Namely, in (\ref{martingale1}) the Poisson process increments $\mathrm{d}\nu_{\ell,t}$'s   are compensated by their conditional expectation \eqref{Poisson2}, so that the expectation value of the increments of $ \mu_{t} $ conditional on the values of $\bm{\psi}_t$, $\bm{\psi}_t^\dagger$ vanishes at any time instant $ t $:
\begin{align}
	\operatorname{E}(\mathrm{d}\mu_{t}|\bm{\psi}_{t}, \bm{{\psi}}^\dagger_{t})=0.
	\label{eq:local-martingale}
\end{align}
A local martingale becomes a strict martingale, i.e. satisfies the condition
$\operatorname{E}\mu_{t}=1$ for all $t$, if the integrability condition $ \operatorname{E}\sup_{t}|\mu_{t}|<\infty $ holds. In practice, we expect $ \mu_t $ to be a strict martingale if all $\Gamma_{\ell,t}$'s are bounded functions of $ t $ during the evolution horizon. We take for granted this physically reasonable condition, and therefore   that the process  $\mu_t $ is a strict martingale.

The last step in the proof of the unraveling via the influence martingale is to compute the differential of  the expectation value (\ref{interference})  using (\ref{sse}), (\ref{Poisson}), and  (\ref{martingale}).  A straightforward application of stochastic calculus proves that the state operator satisfies the time local master equation \eqref{LGKS}. We report the details of the calculation in Methods. 

The question naturally arises whether the evolution law (\ref{sse}) preserves the squared norm of the stochastic process $  \bm{\psi}_{t}  $, and, as a consequence, justifies the interpretation of $ \bm{\psi}_{t}  $ as state vector of the  system. 
 We verify that the squared norm satisfies the It\^{o} stochastic differential equation
\begin{align}
	\mathrm{d}\left(\left\|\bm{\psi}_{t}\right\|^{2}\right)=	
	\sum_{\ell=1}^{\mathscr{L}} \big{(}\mathrm{d}\nu_{\ell,t}-%(1+\mathscr{g}_{\ell,t})\mathscr{r}_{\ell,t}
\Gamma_{\ell,t}\left\|\operatorname{L}_{\ell}\bm{\psi}_{t}\right\|^{2}\mathrm{d}t \big{)}\left(1-\left\|\bm{\psi}_{t}\right\|^{2}\right).
	\nonumber
\end{align}
Therefore, for arbitrary initial values $\bm{\psi}_0$, $\bm{\psi}_0^\dagger$ the expected value of the squared norm is not preserved unless \eqref{CP} holds true. Nevertheless, the Bloch hyper-sphere (i.e. the manifold $\|\bm{\psi}_t\|^2=1$) is preserved by the dynamics. Thus, we can interpret $ \bm{\psi}_{t} $ as a state vector for any quantum trajectory evolving from physically relevant initial data assigned on the Bloch hyper-sphere. A further useful consequence, is that Bloch hyper-sphere valued solutions of (\ref{sse})  can always be couched into the form of the ratio of the solution of a linear stochastic differential equation divided by its norm. We defer the proof of the claim to Methods.

\paragraph{Remarks.} Some observations are in order regarding  equations (\ref{sse}), (\ref{martingale}). 
The evolution law (\ref{martingale}) of the influence martingale is enslaved to that of the state vector (\ref{sse}): $ \mu_{t} $ exerts no feedback on the stochastic Schr\"odinger equation (\ref{sse}). {Most importantly, the state vector is a Markov process and the influence martingale is also non anticipating.}
This is intuitively pleasing because in any finite dimensional Hilbert space the master equation (\ref{LGKS}) is just a matrix-valued time non-autonomous linear ordinary differential equation. 
{Finally, we emphasize the different nature of the weights $ \Gamma_{\ell,t} $'s and of the positive definite rates $ \mathscr{r}_{\ell,t} $'s. The former ones are theoretical predictions fixed by the microscopic dynamics. The $ \mathscr{r}_{\ell,t} $'s are either inferred from experimental measurement  or, in numerical applications, selected based on integration convenience.  Such arbitrariness reflects the fact that quantum trajectories generated by an unraveling exist only contextually to a setup or in the language of \cite{Wis1996} are ``subjectively real''.}

\paragraph{Interpretation.} Girsanov formula (see e.g. \cite{Kle2005}) states that, when the  martingale process $   \mu_{t} $ in (\ref{interference}) is positive definite, it specifies a change of probability measure. The influence martingale can, however, take negative values when the $\Gamma_{\ell,t}$'s do so. {As $ \mu_{t} $ is non-anticipating, at any time $ t $ it is always possible to represent it as the difference of two positive definite and correlated processes 
\begin{align}
	\mu_{t}^{(\pm)}=\max(0,\pm \mu_{t}).
	\nonumber
\end{align}
The immediate consequence is that we can couch (\ref{interference}) into the form
\begin{align}
	\bm{\rho}_{t}=\operatorname{E}\left(\mu_{t}^{(+)}\psi_{t}\bm{\psi}_{t}^{\dagger}-\mu_{t}^{(-)}\psi_{t}\bm{\psi}_{t}^{\dagger}\right)
	\label{WPr}
\end{align}
Using the explicit expression of the influence martingale and of the state vector in terms of the fundamental solution of the linear dynamics  (Methods) it is straightforward to verify that for any realization of the Poisson process the argument of the expectation value is the difference of two completely positive dynamical maps. 
We thus recognize that from the mathematical point of view, the need to introduce the influence martingale naturally stems from the general result in linear operator algebra known as the Wittstock-Paulsen decomposition \cite{Pau2002} stating that any completely bounded map is always amenable to the difference of two completely positive maps.}

If one insists on the change of probability measure interpretation, negative values of the influence martingale would imply that some realizations of quantum trajectories in the mathematical path-space should be weighed by a ``negative probability''. 
Exactly for the reasons put forward by Feynman in \cite{Fey1987}, even such interpretation does not pose any logical difficulty {when the initial state operator belongs to the compatibility domain of operators whose positivity is preserved by the evolution \cite{JoShSu2004,ShSu2005}}. 
Negative values of the influence martingale  only contribute as an intermediate step to the Monte Carlo evaluation of the state operator. In other words, they do not specify ``the final probability of verifiable physical events'' \cite{Fey1987}.  {We refer to \cite{AbBr2011,AbBr2014,BlGu2021} for a mathematically rigorous operational definition of negative probabilities in quantum mechanics recently developed starting from Feynman's  argument.} 
More interesting is the identification of the physical origin of negative values of the influence martingale. Again in \cite{Fey1987} (pag. 246-48) Feynman analyzes a double-slit experiment. 
He notices that probability distributions of certain verifiable events and predicted by Quantum Mechanics exhibit interference patterns. Feynman then shows that interference patterns can be equivalently computed by means of classical, i.e. arithmetic, averages extended to events, not directly verifiable, which are individually weighed by a negative probability. 

It seems therefore to us that it is precisely the need to take into account interference between quantum trajectories in path-space that furnishes the most plausible explanation for the origin of the influence martingale. 
The path integral formulation of quantum mechanics \cite{FeHi2010} offers a further mathematical argument in the same direction. The  influence martingale is a ``shadow on the cave wall'' of the  interference between paths  in the Feynman path integral describing the exact system-environment unitary dynamics. In this sense, $ \mu_t $ plays a role reminiscent of the influence functional introduced by Feynman and Vernon in \cite{FeVe1963}: it encodes information about the coupling to the environment.

\subsection{Examples} {In low dimensional Hilbert spaces, algorithms based on quantum trajectories are not expected to bring any numerical efficiency advantage with respect to direct integration of the master equation. Thus the purpose of the examples is only to highlight how a non-anticipating unraveling reproduces physical phenomena such as quantum revivals usually attributed to memory and prescience effects. In all the examples we set for convenience $ \mathscr{r}_{\ell,t}=|\Gamma_{\ell,t}| $ in \eqref{Poisson}.}

\paragraph{Photonic band gap} \label{ex:pbg} The master equation in Dirac's interaction picture of a two level atom in a photonic band gap \cite{JoWa1990,JoQu1994}  is
\begin{align}
	\bm{\dot{\rho}}_{t}=
	\frac{S_{t}}{2\,\imath}\left[\sigma_{+}\sigma_{-}\,,\bm{\rho}_{t}\right]
	+
	\Gamma_{t}\left(\left[\sigma_{-}\bm{\rho}_{t}\,,\sigma_{+}\right]+\left[\sigma_{-}\,,\bm{\rho}_{t}\sigma_{+}\right]\right)
	\label{pbg:me}
\end{align}
where $\sigma_{\pm}=(\sigma_{1}\pm\imath\,\sigma_{2})/2$ and $ \left\{ \sigma_{i} \right\}_{i=1}^{3} $ are Pauli matrices.  The time dependent functions $ S_{t} $ and $ \Gamma_{t} $ are respectively the Lamb shift and the Lindblad weight factor. 
Negative values of $ \Gamma _{t}$ also imply a violation of the Kossakowski conditions (\cite{Kos1972} see also e.g, \cite{RiHu2012,BrLaPiVa2016}) a weaker form of positivity that might be imposed on $\bm{\rho}_t$  \cite{ChMa2014}. This fact renders the unraveling of (\ref{pbg:me}) in quantum trajectories particularly probing.  

In order to explore a genuine strong system-environment coupling, we proceed as in \cite{PiMaHaSu2008}. Let $\bm{g}$ and $\bm{e}$ be respectively the ground and excited state of $\sigma_{+}\sigma_{-}$.  Using the solution of  the off-diagonal matrix element  $ \bm{e}^\dagger \bm{\rho}_t \bm{g}=c_t\,\bm{e}^\dagger \bm{\rho}_0 \bm{g}$ of a qubit in a photonic band gap (equation (2.21) of \cite{JoQu1994} with $\beta=-\delta$), we determine the Lamb shift $S_t$ and weight factor $\Gamma_t$ by \cite{BrPe2002}
\begin{equation}\label{eq:johnrate}
    S_t=-2\,\textrm{Im}\frac{\dot{c}_t}{c_t},\quad \Gamma_t=-2\,\textrm{Re}\frac{\dot{c}_t}{c_t}
\end{equation}
We also translate the origin of time to $ t\approx 1.4 $ so that $ S_{t} $ and $ \Gamma_{t} $ vanish at time origin. 

\begin{figure}%[H]
    \begin{tabular}{cc}
    \includegraphics[scale=0.5]{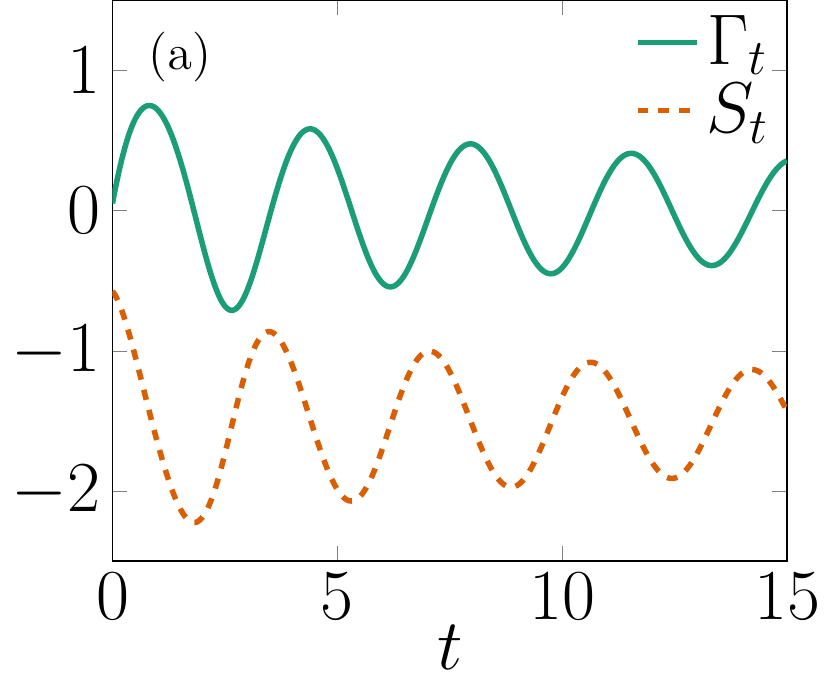}&
    \includegraphics[scale=0.5]{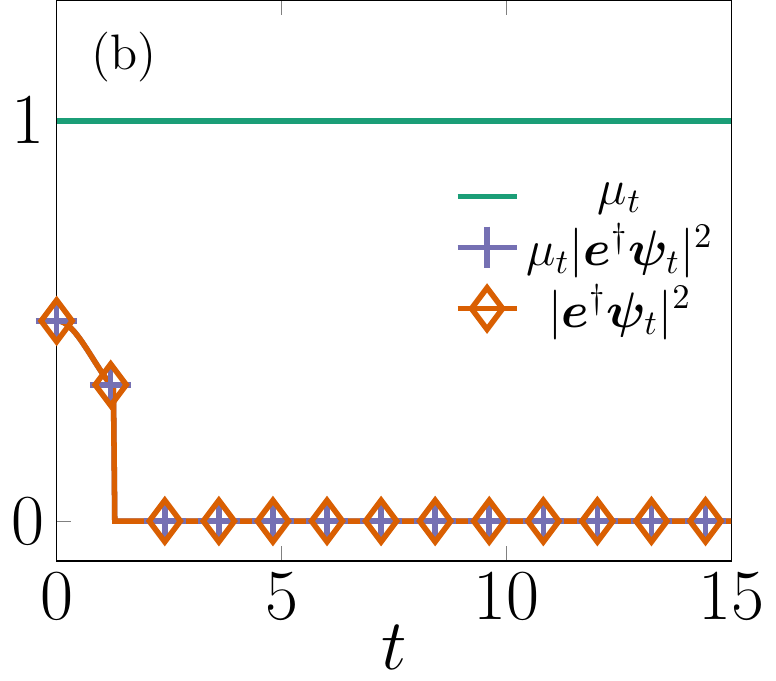}\\
    \includegraphics[scale=0.5]{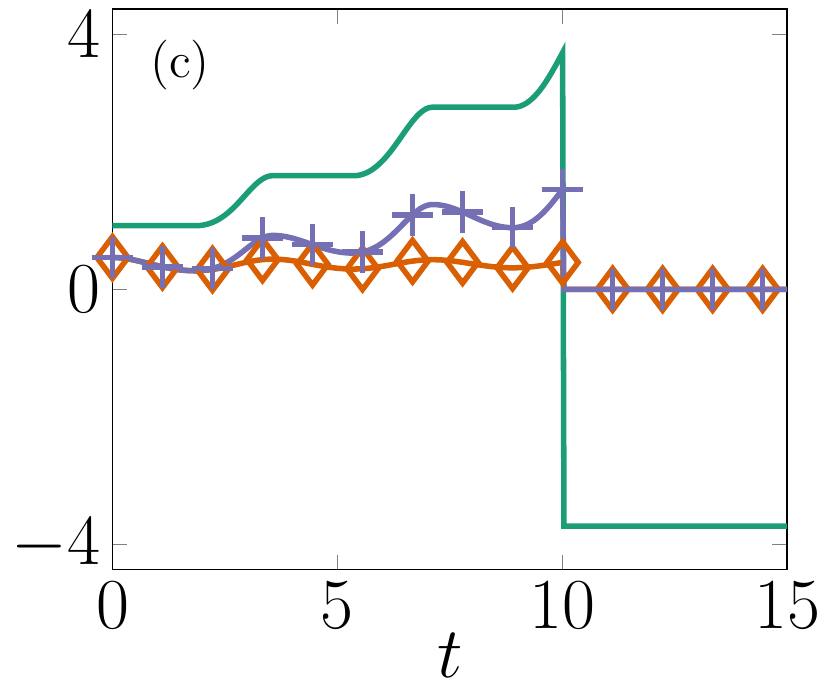}&
    \includegraphics[scale=0.5]{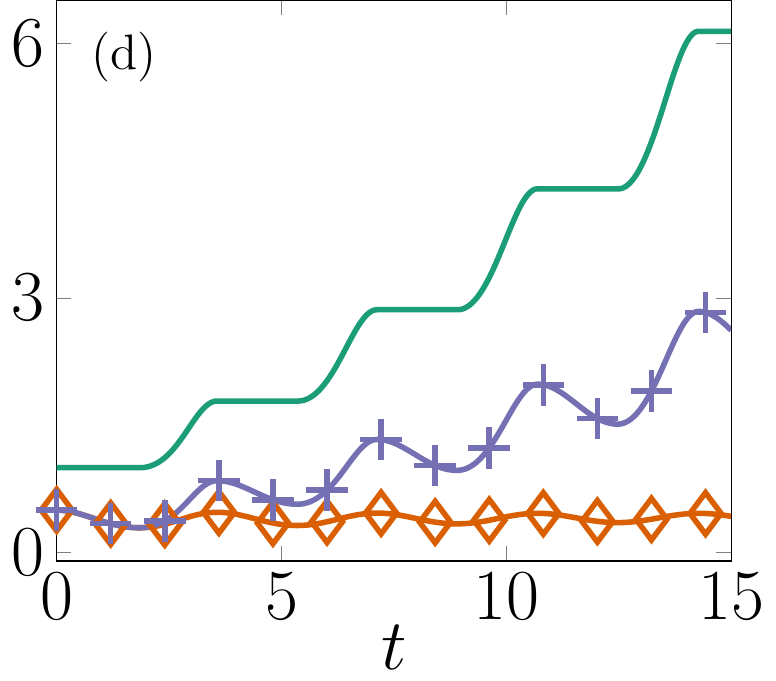}
    \end{tabular}
	\caption{(a) Time dependence of the weight function $\Gamma_t$ (full line) and the Lamb shift $S_t$ (dashed), defined as in \eqref{eq:johnrate}. (b), (c), (d) Different realizations of the stochastic evolution displaying $\left\| \bm{e}^\dagger\bm{\psi}_{t}\right\|^{2}$ (diamonds), $ \mu_{t}\left\|\bm{\psi}_{t}\right\|^{2}  $ (crosses) and  $ \mu_{t} $ (full line). The initial data are $ \bm{\psi}_{0}=(\bm{e}+\bm{g})/\sqrt{2} $ where  $ \bm{g} $ and $ \bm{e} $  are respectively the ground and excited states of $ \operatorname{H}=\sigma_{+}\,\sigma_{-} $.} 
	\label{fig:martingale}
\end{figure}

In Fig.~\ref{fig:martingale} (a) we show the time dependence of the Lindblad weight factor $ \Gamma_{t} $ and the Lamb shift $ S_{t} $. In Fig.~\ref{fig:martingale} (c-d) we show typical realizations of $\mu_t$.
In particular, Fig.~\ref{fig:martingale} (d) exhibits the exponential growth of $\mu_t$ when $\Gamma_t$ is negative and in the absence of jumps.
On the other hand, when $ \Gamma_{t} $ is positive $ \mu_{t} $ is constant in between jumps. Finally in Fig.~\ref{fig:martingale} (c) we show a realization of $\mu_t$ taking negative values in consequence of a quantum jump.

\begin{figure}%[H]
	\centering
 {\includegraphics[scale=1]{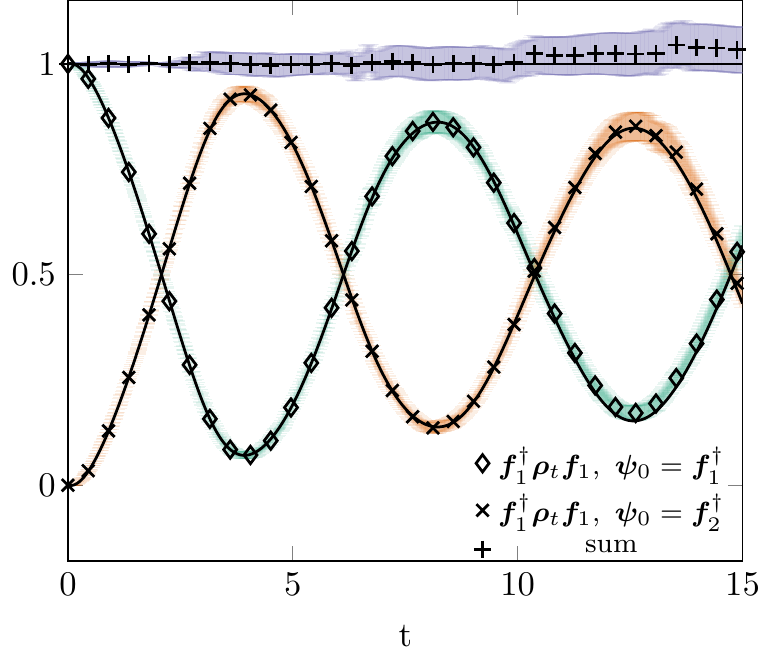}}
	\caption{Monte Carlo averages versus master equation predictions for \eqref{pbg:me}. The scatter plots show $\bm{f}^\dagger_1\bm{\rho}_t\bm{f}_1$, with $\bm{f}_1=(\bm{e}+\bm{g})/\sqrt{2}$, with initial states $\bm{\psi}_0=\bm{f}_1$ (diamonds) and $\bm{\psi}_0=\bm{f}_2=(\bm{e}-\bm{g})/\sqrt{2}$ (x's) and their sum (crosses) obtained from $10^4$ realizations. The continuous lines show the solutions obtained from directly integrating the master equation. The shaded area is determined by two times the square root of the ensemble variance of the indicator of interest. The starting time mesh for the adaptive code is $\mathrm{d}t= 0.03. $}
	\label{fig:JoQu2}
\end{figure}

Fig. \ref{fig:JoQu2} reports the result of our numerical integration for distinct values of the initial data. We generate the quantum trajectories by mapping \eqref{sse} into a linear equation as described in Methods. 
The (black) full lines always denote  predictions from the master equation. Monte Carlo averages are over ensembles of $10^4$ realizations. We theoretically estimate errors with twice the square root of the ensemble variance of the indicator of interest. 
In all cases, Monte Carlo averages and master equation predictions are well within fluctuation induced errors.
The occurrence of "quantum revivals"  can be also quantitatively substantiated observing that the measure of system-environment information flow introduced in \cite{BrLaPi2009} is simply related to $ \Gamma_{t} $ for this model \cite{BrLaPi2009, BrLaPiVa2016}. Namely, equation (66) of \cite{BrLaPiVa2016} relates the direction of the information flow in the model to the sign of the time derivative of $ |\Gamma_{t}| $.

We perform the numerical integration using the Tsitouras $5/4$ Runge-Kutta method automatically switching for stiffness detection to a  $4$-th order A-stable Rosenbrock. The Julia code  is offered ready for use in the  "\emph{DifferentialEquation.jl}" open source suite \cite{RaNi2017}.

\begin{figure}%[H]
	\centering
	\begin{tikzpicture}[scale=1]
		\node at (0,0) {\includegraphics{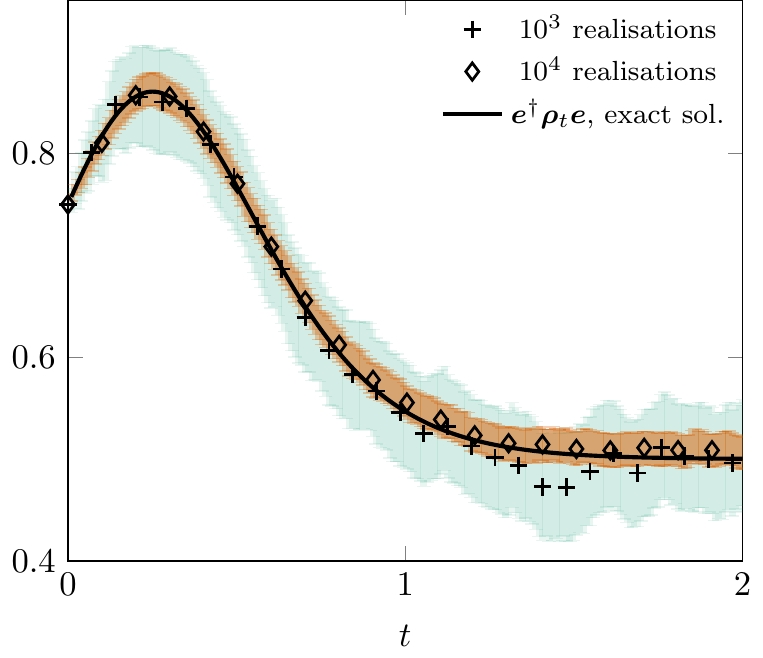}};
	\end{tikzpicture}
	\caption{Monte Carlo averages versus master equation predictions for the energy level populations of the random unitary model \eqref{eq:randomU}. The {Lindblad weights} are $\Gamma_{1,t}=-0.5+2\tanh(\sqrt{2}\,t)$, $\Gamma_{2,t}=-1+2\tanh(\sqrt{3}\,t)$ and $\Gamma_{3,t}=-0.8+2\tanh(\sqrt{5}\,t)$.  The initial condition of the qubit is $\bm{\psi}_0=\frac{\sqrt{3}}{2}\bm{e}+\frac{1}{2}\bm{g}$ where $\bm{g},\, \bm{e}$ are the ground and excited state of $\sigma_3$, respectively. The black full line gives the prediction by solving the master equation  (\ref{eq:randomU}). The crosses show the stochastic average after $10^3$ realizations and the (light) green shaded area displays fluctuation related errors estimated as in Fig.~\ref{fig:JoQu2}. Similarly, the diamonds show the ensemble average after $10^4$ realizations and the (dark) brown shaded area the estimated error.}
	\label{fig:randomU_NoP}
\end{figure}

\paragraph{Qubit model with "controllable positivity".} The master equation
\begin{equation}\label{eq:randomU}
    \bm{\dot{\rho}}_t= \sum_{\ell=1}^{3} \Gamma_{\ell,t}(\sigma_{\ell}\bm{\rho}_t\sigma_{\ell} -\bm{\rho}_t)
\end{equation}
provides a mathematical model of an all-optical setup exhibiting controllable transitions from positive-divisible to non-positive divisible evolution laws \cite{BeCuOrMo2015}.
We focus on the case when the Lindblad weights are of the form $(\ell=1,\,2,\,3)$
\begin{align}
	\Gamma_{\ell,t}=-a_{\ell}+b_{\ell}\tanh(c_{\ell}\,t)
	\nonumber
\end{align}
We choose  $\,a_{\ell},\,b_{\ell},\, c_{\ell}>0$ such that all {Lindblad weights} are negative during a finite time interval around $t=0$. As a consequence, not only the Kossakowski conditions are violated but the rate operator \cite{CaSmBa2017} has negative eigenvalues for the initial conditions we consider.  
Under these hypotheses, and to the best of our knowledge, only the influence martingale permits to unravel  the master equation in quantum trajectories taking values in the Hilbert space of the system.
Fig. \ref{fig:randomU_NoP} shows the evolution in time of $ \bm{e}^{\dagger}\bm{\rho}_{t}\bm{e} $ where $ \bm{e} $ is the excited state of the Pauli matrix $\sigma_3$, i.e. it has eigenvalue 1.
The full (black) line is the master equation prediction, the crosses show the Monte Carlo average \eqref{interference} for $10^3$ realizations and the diamonds for $10^4$. The  light green shaded area shows twice the square root of the variance for $10^3$ realizations and the darker brown for $10^4$. In agreement with the theory, increasing the number of realizations in the ensemble evinces convergence.

\begin{figure}%[H]
	\centering
	\begin{tikzpicture}[scale=1]
		\node at (0,0) {\includegraphics{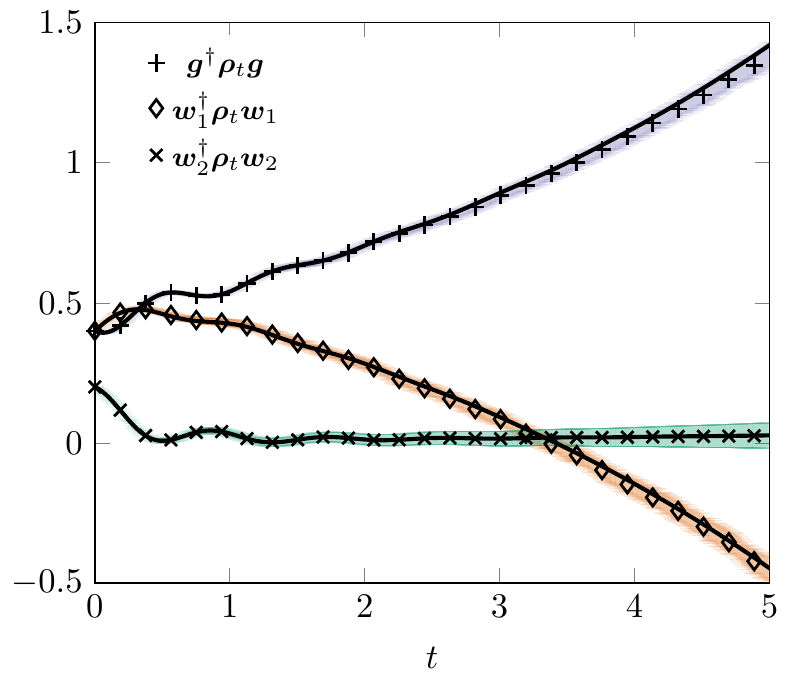}};
	\end{tikzpicture}
	\caption{Redfield equation model, with parameters $\gamma_1=1$, $\gamma_2=4$, $\alpha=3$ and $\kappa=1$. The initial condition is $\bm\psi_0=\sqrt{0.2}\,\bm{w}_1+\sqrt{0.1}\,\bm{w}_2+\sqrt{0.7}\,\bm{g}$ where $\bm{g}$ is the ground state, $\bm{w}_1=L_{1}^{\dagger}\,\bm{g}$ and $\bm{w}_2=L_{2}^{\dagger}\,\bm{g}$. The crosses, diamonds and x's show the result from Monte Carlo simulations after $10^4$ realizations. The full lines are master equation  predictions and the shaded regions have the same meaning as in Fig.~\ref{fig:JoQu2} . The time mesh is $\mathrm{d}t=0.0125$. The diamonds show that $\bm{\rho}_t$ is non positive definite for $t>3$}
	\label{fig:redfield}
\end{figure}

\paragraph{Redfield Equation} As a further example of how the influence martingale associates a quantum trajectory picture even to master equations with non positive definite solutions, we consider a Redfield equation model. We call Redfield a master equation of the form (\ref{LGKS}) obtained from an exact system-environment dynamics by implementing the Born-Markov approximation without a rotating wave approximation. Redfield equations are phenomenologically known to give accurate descriptions of the system dynamics at arbitrary system-environment coupling although they do not guarantee a positive time evolution of the state operator see e.g. \cite{MoLi2020,BrPe2002, HaSt2020}.
In Methods we outline the derivation of a Redfield equation from the exact dynamics of two non-interacting qubits in contact with a boson environment \cite{Dav2020}.
The result is an equation of the form (\ref{LGKS}) with only two Lindblad operators $ \left\{ \operatorname{L}_{\ell} \right\}_{\ell=1}^{2} $ satisfying the commutation 
relations
\begin{align}
	[\operatorname{L}_{\ell}\,,\operatorname{L}_{\mathscr{k}}^\dagger]=\delta_{\ell\,\mathscr{k}}\,,
	\hspace{1.0cm}[\operatorname{L}_{\ell}\,,\operatorname{L}_{\mathscr{k}}]=0.
	\nonumber
\end{align}
The corresponding weights are time independent
\begin{align}
	\Gamma_{\ell,t}=\lambda_{\ell}=\frac{\gamma_1+\gamma_2+(-1)^{\ell}\sqrt{2}\sqrt{\gamma_1^2+\gamma_2^2+2\,\kappa^2}}{4}
\label{Redfield:weight}
\end{align}
where  $ \gamma_{1} $, $ \gamma_{2} $, $ \kappa $ are real numbers. We refer to Methods for the explicit expressions of $ \operatorname{H} $ and $ \left\{ \operatorname{L}_{\ell} \right\}_{\ell=1}^{2} $. Inspection of (\ref{Redfield:weight}) shows that $  \lambda_{1}$ is negative definite. Fig. \ref{fig:redfield} shows how the influence martingale reproduces the predictions of the Redfield equation.

\subsection{Applications}

{\paragraph{Simulating Large Quantum Systems} 
was the motivation of the authors of \cite{DaCaMo1992} for unraveling a master equation or, in their words, for ``Monte Carlo wave-function methods''. The reason is the following.
Given a $ \mathscr{N} $-state system numerical integration of the master equation requires to store $ O(\mathscr{N}^{2}) $ real numbers so that the computing time typically scales as $ O(\mathscr{N}^{4}) $. 
Unraveling the state operator as the average over state vectors generated by a Markov process requires to store only $ O(2\,\mathscr{N}) $ real numbers for each realization. This means that the computing time scales as $ O(\mathscr{N}^{2}\,\times\,\mathscr{M}) $ where $ \mathscr{M} $ is the number of realizations or just $ O(\mathscr{N}^{2}) $ on a parallel processor \cite{WiMi2009}. Thus in high dimensional Hilbert spaces, also due to the existence of efficient numerical algorithms for stochastic differential equations \cite{PlBrLi2010}, ensemble averages are then expected to offer a real advantage for numerical computation \cite{Wis1996,BrPe2002,WiMi2009}. It is worth noticing that the reasoning is very similar to that motivating the use of Lagrangian in place of Eulerian numerical methods in the context of classical hydrodynamics see e.g. \cite{Fri1995}.
}

{
To exhibit that the argument applies to the influence martingale, we compare computing times versus the dimension of the Hilbert space using  QuTiP \cite{JoNaNo2013}, a standard numerical toolbox for open quantum systems. Specifically we take a chain of $ \mathscr{N} $ coupled qubits and directly integrate the master equation using QuTiP.
Next, we perform the same calculation using the QuTiP package for Monte Carlo wave-function methods combined with the implementation of the influence martingale. To do so we choose the rates in (\ref{Poisson}) as we detail below.
}

{ In (\ref{LGKS}) we take $ \mathscr{L}=2\,\mathscr{N} $ where the Lindblad operator
 $ \operatorname{L}_{\ell}$ for $ \ell=1,\dots,\mathscr{N} $  is the tensor product of the lowering operator of the $ \ell $-th qubit with the identity acting on the Hilbert space of the remaining qubits. Thus we set
 \begin{align}
	\operatorname{L}_{\ell}=\sigma^{(\ell)}_{-}, \hspace{0.5cm}\operatorname{L}_{\ell+\mathscr{N}}=\sigma^{(\ell)}_{+} \hspace{0.5cm}\ell=1,\dots,\mathscr{N}  
	\nonumber
\end{align}
The Hamiltonian is
\begin{align}\label{eq:ham_chain}
	\operatorname{H}=\sum_{\ell=1}^{\mathscr{N}} \sigma_{+}^{(\ell)}\sigma^{(\ell)}_{-}+ \lambda\sum_{\ell=1}^{\mathscr{N}-1}( \sigma^{(\ell)}_{+}  \sigma^{(\ell+1)}_{-}+ \sigma^{(\ell+1)}_{+}  \sigma^{(\ell)}_{-})
\end{align}
For the sake of simplicity, we assume that the Lindblad weights in (\ref{LGKS}) are all strictly positive definite with the only exception of $\Gamma_{1,t} $ which can take negative values.
For any $ \ell $  different from $ 1 $ and $ \mathscr{N}+1 $ we choose the rates of the Poisson processes \eqref{Poisson}
\begin{align}
	\mathscr{r}_{\ell,t}=\Gamma_{\ell,t}\,>\,0\hspace{1.0cm}\ell\neq 1,\mathscr{N}+1 
	\nonumber
\end{align}
Next, we use the fact that given a collection of scalars $  \mathscr{w}_{\ell}  $ on the real axis, it is always possible to find an equal number of positive definite scalars $ \mathscr{r}_{\ell} $ and a negative definite real $ c $ such that the set of equations
\begin{align}
	\mathscr{w}_{\ell}=\mathscr{r}_{\ell} +c
	\nonumber
\end{align} 
is satisfied. We therefore set for any $ t $
\begin{align}
	& 	\Gamma_{1,t}=\mathscr{r}_{1,t}+c_{t}
	\nonumber\\
	&\Gamma_{1+\mathscr{N},t}=\mathscr{r}_{1+\mathscr{N},t}+c_{t}
	\nonumber
\end{align}
where now $ \mathscr{r}_{1,t} $ and $ \mathscr{r}_{1+\mathscr{N},t} $ specify the rates of the Poisson processes $ \mathrm{d}\nu_{1,t} $ and $ \mathrm{d}\nu_{1+\mathscr{N},t} $. As the Lindblad operators satisfy
\begin{align}
		\sigma^{(1)}_+\sigma^{(1)}_-+\sigma^{(1)}_-\sigma^{(1)}_+=\operatorname{1}_{\mathcal{H}}
	\nonumber
\end{align}
on the Bloch hyper-sphere ($\|\bm{\psi}_t\|^2=1$) we arrive  to the identity for the drift terms in \eqref{sse}
\begin{align}
	& 	\Gamma_{1,t}\frac{\sigma^{(1)}_{+}\sigma^{(1)}_{-}-\|\sigma^{(1)}_{-}\bm{\psi}_t\|^2}{2}\bm{\psi}_t+
	\Gamma_{1+\mathscr{N},t}\frac{\sigma^{(1)}_{-}\sigma^{(1)}_{+}-\|\sigma^{(1)}_{+}\bm{\psi}_t\|^2}{2}\bm{\psi}_{t}
	\nonumber\\
	&= 	\mathscr{r}_{1,t}\frac{\sigma^{(1)}_{+}\sigma^{(1)}_{-}-\|\sigma^{(1)}_{-}\bm{\psi}_t\|^2}{2}\bm{\psi}_t+
	\mathscr{r}_{1+\mathscr{N},t}\frac{\sigma^{(1)}_{-}\sigma^{(1)}_{+}-\|\sigma^{(1)}_{+}\bm{\psi}_t\|^2}{2}\bm{\psi}_{t}
	\nonumber
\end{align}
We conclude that the state operator evolves on the Bloch hyper-sphere according to the same stochastic Schr\"odinger equation of \cite{DaCaMo1992}. We can therefore directly use the Monte Carlo wave function package of QuTiP for computing the evolution of the state vector. Using this information, we compute the influence martingale for each trajectory and finally the state operator. We refer to Methods for further details.}

\begin{figure}
    \centering
    \includegraphics{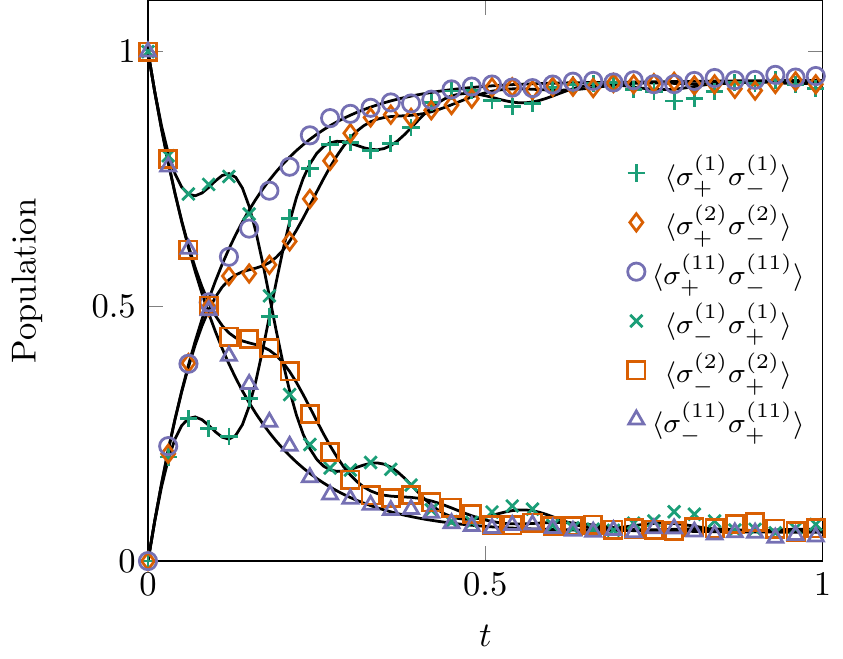}
    \caption{Population of sites 1, 4 and 11 for the qubit chain \eqref{eq:ham_chain} with $L=11$. The marks show the result of the stochastic evolution and the full black lines the result of numerically integrating the master equation. The weights are $\Gamma_{\ell,t} =\gamma$, $\Gamma_{\ell+N,t} =\delta $ for $\ell=1,\hdots N$, $\Gamma_{1+N,t}=\delta$ for all $t$, $\Gamma_{1,t}=\Gamma_{1,t}=\gamma - 12 \exp(-2 t^3)\sin^2(15t)$ with $\gamma = (1/0.129) (1 + 0.063) $ and $\delta = (1/0.129) 0.063 $  and $\lambda = 10$.}
    \label{fig:population}
\end{figure}

\begin{figure}
    \centering
    \includegraphics{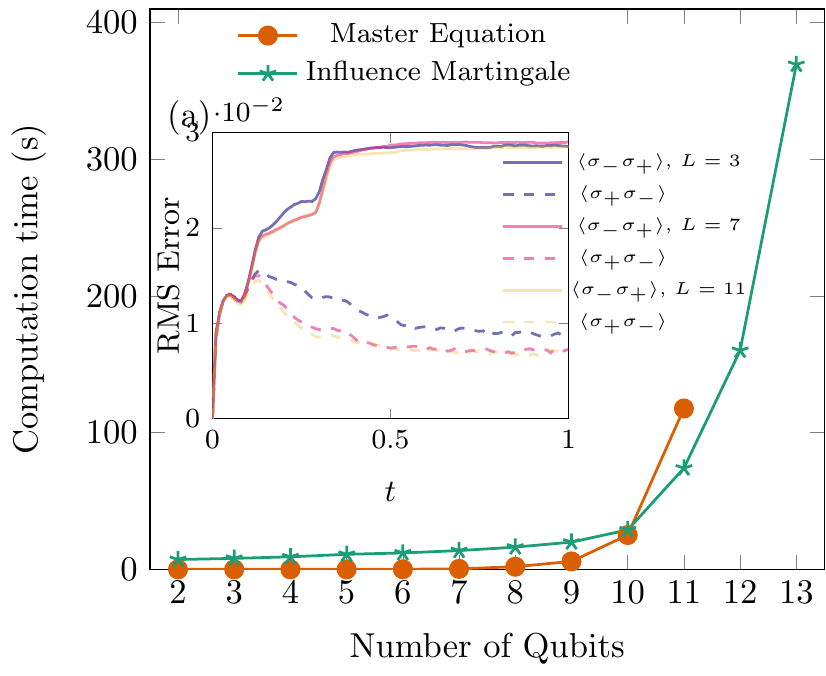}
    \caption{(a) Computation time for both the Master Equation and Influence Martingale method as a function of the amount of qubits in the chain. For the stochastic method we generated 1000 realizations. For $ \mathscr{N}>11 $ the use of the master equation becomes unwieldy on our laptops whereas we were able to take further points using the unraveling. (b) The root mean square error of the populations averaged over all individual sites.}
    \label{fig:Times}
\end{figure}

{Figure \ref{fig:population} shows the populations of several sites for $\mathscr{N}=11$. The computation time is shown in Figure \ref{fig:Times} (a). At $ \mathscr{N}=10 $ we observe a cross-over of the computation time curves. After $ \mathscr{N}=11 $ the influence martingale based algorithm becomes more efficient without applying any adapted optimization. From 12 qubits on, direct integration of the master equation becomes unwieldy (Apple M1 CPU). With the influence martingale, even 13 coupled qubits take only a few minutes of computation time. Most of the computation time is due to generating the trajectories. The actual computation of the martingale takes $0.19$s for $\mathscr{N}=2$ and $0.42$s for $\mathscr{N}=11$. The number of realisations is always $10^3$. Figure \ref{fig:Times} (b) shows the root mean square error of site occupations averaged over all sites. The error stays approximately constant when $\mathscr{N}$ increases.
}

\paragraph{Photo-current oscillations}
{In experimental quantum optics, photo-current is usually defined as the average number of detection events per unit of time.
The stochastic differentials (\ref{Poisson}) mathematically describe increments of the photocurrent associated to detection events corresponding to transitions induced by the Lindblad operator $ \operatorname{L}_{\ell} $ in (\ref{LGKS}) \cite{WiMi2009}. }

{The study of time-dependent transport properties of photo-excited undoped super-lattices highlighted the phenomenon of photo-current oscillations \cite{BoGaCuMa1994}. Photo-current oscillations naturally come about when working with the time convolutionless master equation with non positive Lindblad weights. In such a case and under
standard hypotheses (see Methods) it is straightforward to verify that  a system governed by an Hamiltonian $ \operatorname{H}_{\mathfrak{0}} $ when isolated satisfies when the coupled to an environment the energy balance equation
\begin{align}
\mathrm{d}\operatorname{Tr}(\operatorname{H}_{\mathfrak{o}}\bm{\rho}_{t})=
	\frac{ \operatorname{Tr}\left[\operatorname{H}_{\mathfrak{o}}\,,\operatorname{H}_{t}\right]\bm{\rho}_{t}\mathrm{d}t}{\imath}+
	\sum_{\ell=1}^{\mathscr{L}}\epsilon_{\ell}\,\mathrm{d}(\operatorname{E}\mu_{t}\nu_{\ell,t})
	\label{pc:eb}
\end{align}
where the $ \epsilon_{\ell} $ are the energy quanta exchanged in transitions. The expression generalizes the results of \cite{WiMi2009} showing that, as the strength of the system environment coupling increases, the average values of photo-current increments are modulated by the influence martingale in thermodynamic relations. The conclusion is that photo-current oscillations reflect a heat flow both from and to the system.}
 
\section{Discussion} {In this paper we proved the existence of  a non-anticipating unraveling of general time local master equations taking values on the Bloch hyper-sphere of the system. }

In \cite{Wis1996} Wiseman discussed three interpretations of quantum trajectories generated by unravelings. {Here, exactly as in \cite{Wis1996}, the order of listing is absolutely not meant to reflect importance.}
The first interpretation is as mathematical tool to compute the solution of the Lindblad--Gorini--Kossakowski--Sudarshan equation  for high dimensional systems (see e.g. \cite{DaCaMo1992}). The second interpretation of quantum trajectories construes them as {subjectively real}: their existence and features are determined only {contextually} to a given physical setup (see e.g. \cite{WiMi1992,BrPe1995}). And finally, quantum trajectories might be an element of a still missing theory of quantum state reduction \cite{Gis1984,GhRiWe1986,Per2003,BaGh2003}. We believe that the influence martingale yields a significant contribution to, at least, the first two interpretations. 
As a mathematical tool, the unravelling via influence martingale method only involves the use of ordinary stochastic differential equations with Poisson noise. The proof is rigorous and straightforward. As a result, the convergence of Monte Carlo averages is guaranteed by standard results in the theory of stochastic differential equations \cite{PlBrLi2010}. 
A further advantage is that the unraveling does not rely on any hypothesis on the sign of the scalar prefactors, weights, of the Lindblad operators in the master equation. 
{Generalising what was previously established for master equations derived in the weak coupling scaling limit \cite{BrPe2002,WiMi2009}, we here  provide explicit evidence of the advantage of using the influence martingale to integrate a substantially larger class of master equations when the dimension of the Hilbert space is large.  Finally,  we observe that if the purpose of introducing quantum trajectories is limited to numerical applications, resorting to the generation of ostensible statistics as in \cite{GaWi2001} may further speed up calculations. For ostensible statistics trace preservation in (\ref{interference}) holds not pathwise but only on average, Bloch hyper-sphere conservation is thus not required and the influence martingale can be replaced by a simple jump process. }

{Regarding the second interpretation,  the meaning of the influence martingale is that of representing the completely bounded fundamental solution of the universal form of the time local master equation as a statistical average over stochastic realizations of completely bounded maps. General results \cite{Pau2002} in linear operator algebra prove that a completely bounded map can be embedded into a completely positive map. Combining this fact with non-anticipating nature of the unraveling guarantees the measurement interpretation (Methods).   An explicit example of the construction of the embedding is given in \cite{Bre2004}. There the unraveling was only defined in the extended Hilbert space thus requiring the introduction of a ``minimal" dynamics for an auxiliary environment. Here we prove that quantum trajectories can be computed directly on the Bloch hyper-sphere of the system, their occurrence being always consistent with a measurement performed on an environment that does not need to be specified. 
}

{	
In conclusion, the main result of the present paper substantially extends the domain of application of quantum trajectory based methods of  state and dynamical parameter estimation, prediction and retrodiction as recently reviewed in \cite{WeMuKiScRoSi2016}.}

\section{Methods} 

\subsection{Derivation of the Master Equation.}

In order to verify that  the influence martingale representation of the state operator \eqref{interference} generically yields a solution of the master equation (\ref{LGKS}) we evaluate
\begin{equation}\nonumber
    \mathrm{d}\bm\rho_t=\mathrm{d}\operatorname{E}\left(\mu_{t}\bm{\psi}_{t}\bm{\psi}_{t}^{\dagger}\right)
    %=\operatorname{E}\left(\mathrm{d}\left(\mu_{t}\bm{\psi}_{t}\bm{\psi}_{t}^{\dagger}\right)\right)
\end{equation}
along the paths of the stochastic differential equations (\ref{sse}) and (\ref{martingale}).
According to the rules of stochastic calculus, see e.g. \cite{Kle2005}, differentiation commutes with the expectation value operation. We need therefore to apply It\^o lemma to evaluate
\begin{align*}
&\mathrm{d}(\mu_t\bm{\psi}_t\bm{\psi}_t^\dagger)=(\mathrm{d}\mu_t)\bm{\psi}_t\bm{\psi}_t^\dagger+\mu_t\mathrm{d}(\bm{\psi}_t\bm{\psi}_t^\dagger)+ (\mathrm{d}\mu_t)\mathrm{d}(\bm{\psi}_t\bm{\psi}_t^\dagger)
\end{align*}
 in the case of Poisson noise. The stochastic differential of the outer product of the state vector with its complex adjoint
\begin{align*}
\mathrm{d}(\bm{\psi}_t\bm{\psi}_t^\dagger)=  (\mathrm{d}\bm{\psi}_t)\bm{\psi}_t^\dagger+\bm{\psi}_t(\mathrm{d}\bm{\psi}_t^\dagger)+(\mathrm{d}\bm{\psi}_t)(\mathrm{d}\bm{\psi}_t^\dagger)
\end{align*}
evaluated along (\ref{sse}) is
\begin{align*}
	&\mathrm{d}(\bm{\psi}_t\bm{\psi}_t^\dagger)=-\imath [\operatorname{H}_t,\bm{\psi}_t\bm{\psi}_t^\dagger] \mathrm{d}t
	\\
	&-\sum_{\ell=1}^{\mathscr{L}}\Gamma_{\ell,t} \left(
	\frac{\operatorname{L}_\ell^\dagger \operatorname{L}_\ell \bm{\psi}_t\bm{\psi}_t^\dagger+\bm{\psi}_t\bm{\psi}_t^\dagger\operatorname{L}_\ell^\dagger \operatorname{L}_\ell }{2}-\|\operatorname{L}_\ell\bm{\psi}_t\|^{2}\bm{\psi}_t\bm{\psi}_t^{\dagger}
	\right)
	\mathrm{d}t\\
	&+\sum_{\ell=1}^{\mathscr{L}} \left(\frac{\operatorname{L}_{\ell,t}\bm{\psi}_t\bm{\psi}_t^{\dagger}\operatorname{L}_{\ell,t}^{\dagger} }{\|\operatorname{L}_\ell \bm{\psi}_t\|^2}-\bm{\psi}_t\bm{\psi}_t^\dagger\right)\mathrm{d}\nu_{\ell,t}.
\end{align*}
Combining the above equation with \eqref{martingale} and using the properties of Poisson differentials  (\ref{Poisson}), yields
\begin{align*}
    &\mathrm{d}(\mu_t\bm{\psi}_t\bm{\psi}_t^\dagger)=(\mathrm{d}\mu_t)\bm{\psi}_t\bm{\psi}_t^\dagger-\imath\, \mu_t\,[\operatorname{H}_t,\bm{\psi}_t\bm{\psi}_t^\dagger] \mathrm{d}t\\
    &-\mu_t\sum_{\ell=1}^{\mathscr{L}}\Gamma_{\ell,t}\, \left(
    \frac{\operatorname{L}_\ell^\dagger \operatorname{L}_\ell \bm{\psi}_t\bm{\psi}_t^\dagger+\bm{\psi}_t\bm{\psi}_t^\dagger\operatorname{L}_\ell^\dagger \operatorname{L}_\ell }{2}
    -\|\operatorname{L}_\ell\bm{\psi}_t\|^{2}\bm{\psi}_t\bm{\psi}_t^\dagger 
    \right)\mathrm{d}t\\
    &+\mu_t\sum_{\ell=1}^{\mathscr{L}}\frac{\Gamma_{\ell,t}}{\mathscr{r}_{\ell,t}}\left(\frac{\operatorname{L}_{\ell}\bm{\psi}_t\bm{\psi}_t^{\dagger}\operatorname{L}_{\ell}^{\dagger}}{\|\operatorname{L}_\ell \bm{\psi}_t\|^2} -\bm{\psi}_t\bm{\psi}_t^\dagger\right)\mathrm{d}\nu_{\ell,t}.
\end{align*}

Combining the telescopic property of conditional expectations see e.g. \cite{Kle2005} and the martingale property (\ref{eq:local-martingale}) yields
\begin{align}
	\operatorname{E}(\mathrm{d}\mu_t\bm{\psi}_t\bm{\psi}_t^\dagger)=0
	\nonumber
\end{align}
The remaining terms under expectation reduce to
\begin{align*}
	   &\bm{\dot{\rho}}_{t}=-\imath \,[\operatorname{H}_t,\bm{\rho}_{t}] \\
	&-\sum_{\ell=1}^{\mathscr{L}}\Gamma_{\ell,t}\, \left(\frac{\operatorname{L}_\ell^\dagger \operatorname{L}_\ell\bm{\rho}_{t}+\bm{\rho}_{t}\operatorname{L}_\ell^\dagger \operatorname{L}_\ell}{2}-\operatorname{E}\mu_{t}\|\operatorname{L}_\ell\bm{\psi}_t\|^2\bm{\psi}_{t}\bm{\psi}_{t}^{\dagger} \right)\\
	&+\sum_{\ell=1}^{\mathscr{L}}\Gamma_{\ell,t}\operatorname{E}\mu_t\left(\frac{\operatorname{L}_{\ell}\bm{\psi}_t\bm{\psi}_t^{\dagger}\operatorname{L}_{\ell}^{\dagger}}{\|\operatorname{L}_\ell \bm{\psi}_t\|^2} -\bm{\psi}_t\bm{\psi}_t^\dagger\right) \|\operatorname{L}_\ell \bm{\psi}_t\|^2.
	\nonumber
\end{align*}
Straightforward algebra then allows us to recover (\ref{LGKS}).

\subsection{Linear Stochastic Differential Equation.}
The It\^{o} stochastic differential equation \eqref{sse} preserves the Bloch hyper-sphere. On the hyper-sphere, we can look for solutions of  the stochastic Schr\"odinger equation (\ref{sse}) of the form
\begin{align}
	\bm{\psi}_{t}=\frac{\bm{\varphi}_{t}}{\left\|\bm{\varphi}_{t}\right\|}
	\label{map}
\end{align}
The change of variables (\ref{map}) maps (\ref{sse}) into the linear problem 
\begin{align}
&	\mathrm{d}\bm{\varphi}_{t}=-\imath\,\operatorname{H}_t\bm{\varphi}_{t}
	\nonumber\\
&
-\sum_{\ell=1}^{\mathscr{L}}\left (\frac{\Gamma_{\ell,t}}{2}
	\operatorname{L}_{\ell}^{\dagger}\operatorname{L}_{\ell}\mathrm{d}t
	-\mathrm{d}\nu_{\ell,t}\left(\operatorname{L}_{\ell}-\operatorname{1}\right)\right )\bm{\varphi}_{t}
	\label{Methods:lsde}
\end{align}  
 Once we know $ \bm{\varphi}_{t} $, we can use (\ref{map}) to determine the state vector and the influence martingale. In particular,  the influence martingale always admits the factorization
 \begin{align}
 	\mu_{t}=\exp\left (\int_{0}^{t}\mathrm{d}s\,\sum_{\ell=1}^{\mathscr{L}}\left (\mathscr{r}_{\ell,s}-\Gamma_{\ell,s}\right )\left\|\operatorname{L}_{\ell}\bm{\psi}_{s}\right\|^{2}\right )\,\tilde{\mu}_{t}
 	\label{factor}
 \end{align}
where $ \tilde{\mu}_{t} $ is a pure jump process. This factorization is of use in numerical implementations.
From (\ref{factor}) we readily see that negative values of the $ \Gamma_{\ell,t} $'s exponentially enhance the contribution of the realization of the state vector to the expectation value.

\subsection{Derivation of the stochastic Wittstock--Paulsen decomposition}
{
\paragraph{Explicit expression of the probability measure.} We describe sequences of detection events by first supposing that a fixed number $ n $ of jumps occurs in the time interval $ (\ti,t] $. Next, we suppose that a jump of type  $ \ell_{i} $ with $ \ell_{i} $ taking values in $ \left\{ 1,2,\dots,\mathscr{L} \right\} $ occurs at time at $ s_{i} $ satisfying for $ i=1,\dots,n $ the chain of inequalities
\begin{align}
	t\,>\,s_{n}\,>\,\dots\,>\,s_{i}\,>\dots\,>\,s_{\mathfrak{1}}\,>\,s_{\mathfrak{0}}=\ti 
	\label{simplex}
\end{align}
An arbitrary sequence of detection events $ \omega $ thus corresponds to a $ 2 \,n $-tuple $ \left\{  \ell_{i},s_{i} \right\}_{i=1}^{n}$. The total number $ n $ of jumps  ranges from zero to infinity. We refer to this mathematical description of events as \emph{waiting time representation}. On the Bloch hyper-sphere,  all information about the dynamics of (\ref{sse}) in a time interval $ [s,t) $ during which no jump occurs is encapsulated in the Green function of the linear dynamics (\ref{Methods:lsde})
\begin{align}
	&	\dot{\operatorname{G}}_{t\,s}=-\imath\,\operatorname{H}_{t}\operatorname{G}_{t\,s} -\frac{1}{2}
	\sum_{\ell=1}^{\mathscr{L}}\Gamma_{\ell,t}\operatorname{L}_{\ell}^{\dagger}\operatorname{L}_{\ell}\operatorname{G}_{t\,s} +\delta(t-s)
	\nonumber\\
	&\lim_{t \searrow s}\operatorname{G}_{t\,s} =\operatorname{1}_{\mathcal{H}}
	\nonumber	
\end{align}
Exactly repeating the same steps as in \S~6.1 of \cite{BrPe2002}, we obtain the expression of the state vector conditional upon $ \omega $
\begin{align}
	&	\operatorname{E}(\bm{\psi}_{t}\,|\,\omega=\left\{ \ell_{i},s_{i} \right\}_{i=1}^{n},\bm{\psi}_{\ti}=\bm{z})
	\nonumber\\
	&\hspace{1.0cm}=\frac{\bm{\Lambda}_{t\,\ti}(\left\{ \ell_{i},s_{i} \right\}_{i=1}^{n})\bm{z}}{\left\|\bm{\Lambda}_{t\,\ti}(\left\{ \ell_{i},s_{i} \right\}_{i=1}^{n})\bm{z}\right\|}
	\label{sv}.
\end{align}
By using the identity 
\begin{align*}
    &e^{-\int_{s}^{t}\mathrm{d}u\,\sum_{\ell=1}^{\mathscr{L}}\mathscr{r}_{\ell,u}
		\,\frac{\|\operatorname{L}_{\ell}\operatorname{G}_{u\,s}\bm{z}\|^{2}}{\|\operatorname{G}_{u\,s}\bm{z}\|^{2}}}\\
		&=\left\|\operatorname{G}_{t\,\ti}\bm{z}\right\|^{2} e^{-\int_{s}^{t}\mathrm{d}u\,\sum_{\ell=1}^{\mathscr{L}}(\mathscr{r}_{\ell,u}-\Gamma_{\ell,u})
		\,\frac{\|\operatorname{L}_{\ell}\operatorname{G}_{u\,s}\bm{z}\|^{2}}{\|\operatorname{G}_{u\,s}\bm{z}\|^{2}}}
\end{align*}
we find that the multi-time probability density of the conditioning event is equal to
\begin{align}
	&	\operatorname{p}_{t\,\ti}(\omega=\left\{ \ell_{i},s_{i} \right\}_{i=1}^{n}\,|\,\bm{\psi}_{\ti}=\bm{z})=\delta_{n,0} \frac{\left\|\operatorname{G}_{t\,\ti}\bm{z}\right\|^{2}}{\mathscr{m}_{t\,t_{0}}(\bm{z})}
	\nonumber\\
	&\hspace{0.5cm}+\frac{ (1-\delta_{n,0})\left\|\bm{\Lambda}_{t\,\ti}(\left\{ \ell_{i},s_{i} \right\}_{i=1}^{n})\bm{z}\right\|^{2}}{\mathscr{m}_{t\,s_{n}}(\prod_{j=1}^{n}\operatorname{L}_{\ell_{j}}\operatorname{G}_{s_{j}s_{j-1}}\bm{z})\dots\,\mathscr{m}_{s_{1}\,t_{0}}(\bm{z})}
	\label{edp}
\end{align}
In writing (\ref{sv}), (\ref{edp}) we introduced the tensor valued process
\begin{align}
	&	\bm{\Lambda}_{t\,\ti}(\left\{ \ell_{i},s_{i} \right\}_{i=1}^{n}):=
	\delta_{n,0}\operatorname{G}_{t\ti}
	\nonumber\\
	&\hspace{0.2cm}+(1-\delta_{n,0})	\operatorname{G}_{t\,s_{n}}\prod_{i=1}^{n}\sqrt{\mathscr{r}_{\ell_{i},s_{i}}}\operatorname{L}_{\ell_{i}}\operatorname{G}_{s_{i}s_{i-1}}
	\nonumber
\end{align}
and the scalar
\begin{align}
	\mathscr{m}_{t\,s}(\bm{z}):=e^{\int_{s}^{t}\mathrm{d}u\,\sum_{\ell=1}^{\mathscr{L}}(\mathscr{r}_{\ell,u}-\Gamma_{\ell,u})
		\,\frac{\|\operatorname{L}_{\ell}\operatorname{G}_{u\,s}\bm{z}\|^{2}}{\|\operatorname{G}_{u\,s}\bm{z}\|^{2}}}
	\label{scalar}
\end{align}
which is the value of the influence martingale if no jumps occur in the interval $ [s,t] $.
In order to neaten the notation in (\ref{sv}) and (\ref{edp})  and below, we omit to write the condition
$ \bm{\psi}_{\ti}^{\dagger}=\bm{z}^{\dagger} $ that fully specifies the initial data on the Bloch hyper-sphere.}
{
If we now restrict the attention to the computation of quantum probabilities, we see that  the product of the pure state operator specified by (\ref{sv}) times its probability (\ref{edp}) yields the stochastic dynamical map
\begin{align}
	&	\bm{\Phi}_{t\,\ti}[\omega=\left\{ \ell_{i},s_{i} \right\}_{i=1}^{n}](\bm{z}\bm{z}^{\dagger})=
	\nonumber\\
	&\hspace{0.5cm}\frac{\bm{\Lambda}_{t\,\ti}(\left\{ \ell_{i},s_{i} \right\}_{i=1}^{n})\bm{z}\bm{z}^{\dagger}\bm{\Lambda}_{t\,\ti}^{\dagger}(\left\{ \ell_{i},s_{i} \right\}_{i=1}^{n})}{\mathscr{m}_{t\,s_{n}}(\prod_{j=1}^{n}\operatorname{L}_{\ell_{j}}\operatorname{G}_{s_{j}s_{j-1}}\bm{z})\dots\,\mathscr{m}_{s_{1}\,t_{0}}(\bm{z})}
	\label{sdm}
\end{align}
satisfying by construction the unit trace condition
\begin{align}
	&	\int \mathfrak{m}(\mathrm{d}\omega)\operatorname{Tr}\bm{\Phi}_{t\,\ti}[\omega](\bm{z}\bm{z}^{\dagger}):=
	\nonumber\\
	&	\sum_{n=0}^{\infty}\prod_{i=1}^{n}\sum_{\ell_{i}=1}^{\mathscr{L}}\int_{\ti}^{t}\mathrm{d}s_{i}\operatorname{p}_{t\,\ti}(\omega=\left\{ \ell_{i},s_{i} \right\}_{i=1}^{n})=1.
	\label{normalization}
\end{align}
Thus the dynamical map (\ref{sdm}) takes the form of the generalized operator sum representation \cite{ToKwOhChMa2004,RiHu2012}. It differs from the Choi representation of a completely positive map (see e.g. \cite{Pau2002}) 
in consequence of the non-linear dependence upon the initial state.
}
{
\paragraph{Stochastic Wittstock-Paulsen decomposition.} Multiplying  (\ref{sdm}) by the influence martingale \eqref{factor} occasions the cancellation of any non-linear dependence upon the initial data
\begin{align}
	&	\operatorname{E}(\mu_{t}\bm{\psi}_{t}\bm{\psi}_{t}^{\dagger}|\omega,\bm{\psi}_{\ti}=\bm{z})=
	\nonumber\\
	&\hspace{0.4cm}\bm{\Lambda}_{t\,\ti}^{(+)}(\omega)\bm{z}\bm{z}^{\dagger}\bm{\Lambda}_{t\,\ti}^{(+)\dagger}(\omega)
	-
	\bm{\Lambda}_{t\,\ti}^{(-)}(\omega)\bm{z}\bm{z}^{\dagger}\bm{\Lambda}_{t\,\ti}^{(-)\dagger}(\omega)
\nonumber
\end{align}
with
\begin{align}
&	\bm{\Lambda}_{t\,\ti}^{(\pm)}(\left\{ \ell_{i},s_{i} \right\}_{i=1}^{n})=
\nonumber\\
&	\hspace{0.2cm}\left(\frac{\max(0,\pm\prod_{k=1}^{n}\Gamma_{\ell_{k},s_{k}})}{\prod_{j=1}^{n}\mathscr{r}_{\ell_{j},s_{j}}}\right)^{1/2}
	\bm{\Lambda}_{t\,\ti}(\left\{ \ell_{i},s_{i} \right\}_{i=1}^{n})
	\label{frames}
\end{align}
We thus verify that (\ref{WPr}) is indeed an expectation value over the difference of completely positive stochastic dynamical maps.
}

\subsection{Derivation of the Redfield Equation for a Two-Qubit System}
We consider two non-interacting qubits in contact with an environment consisting of  $ \mathscr{N}\uparrow\infty $ bosonic oscillators
\begin{align*}
&        \hat{\operatorname{H}}=\sum_{\mathscr{i}=1}^{2}\omega^{(\mathscr{i})}\sigma_{+}^{(\mathscr{i})}\sigma_{-}^{(\mathscr{i})}
\nonumber\\
&+\sum_{k=1}^{\mathscr{N}} \left(\epsilon_k \,\operatorname{b}^\dagger_k \operatorname{b}_k+\sum_{\mathscr{i}=1}^{2}g_k(\sigma_{+}^{(\mathscr{i})} \operatorname{b}_k +\operatorname{b}^\dagger_k \sigma_{-}^{(\mathscr{i})})\right)
        \nonumber
\end{align*}
Here $ \sigma_{\pm}^{(\mathscr{i})} $ is the lift to the tensor product Hilbert space of the ladder operators acting on the Hilbert space of individual qubits $ \mathscr{i}=1,2 $.
Tracing out the environment yields the master equation \cite{Dav2020}
\begin{align}
 &   \bm{\dot{\rho}}_t=-\imath\sum_{\mathscr{i},\mathscr{j}=1}^{2}\operatorname{A}_{\mathscr{i}\mathscr{j}}\,[\sigma_{+}^{(\mathscr{j})}\sigma_{-}^{(\mathscr{i})},\bm{\rho}_t]
 \nonumber\\
 &   +\sum_{\mathscr{i},\mathscr{j}=1}^{2}\operatorname{B}_{\mathscr{i}\mathscr{j}}
    \frac{\left[\sigma_{-}^{(\mathscr{i})}\,,\bm{\rho}_t\sigma_{+}^{(\mathscr{j})}\right]+\left[\sigma_{-}^{(\mathscr{i})}\bm{\rho}_t\,,\sigma_{+}^{(\mathscr{j})}\right]}{2}
    \nonumber
\end{align}
where the  $ \operatorname{A}_{\mathscr{i}\mathscr{j}} $'s and $ \operatorname{B}_{\mathscr{i}\mathscr{j}}  $'s are respectively the components of the matrix
\begin{equation}\nonumber
    \operatorname{A}=\begin{bmatrix}
    \alpha & \alpha+\dfrac{\kappa}{2}-\imath\dfrac{\gamma_2-\gamma_1}{4}\\ \alpha+\dfrac{\kappa}{2}-\imath\dfrac{\gamma_1-\gamma_2}{4} & \alpha+\kappa
    \end{bmatrix}
\end{equation}
and 
\begin{equation}
    \operatorname{B}=\frac{1}{2}\begin{bmatrix}\gamma_1&\dfrac{\gamma_1+\gamma_2}{2}-\imath\,\varkappa\\
    \dfrac{\gamma_1+\gamma_2}{2}+\imath\,\kappa&\gamma_2
    \end{bmatrix}
    \nonumber
\end{equation}
The key observation is that the numbers $\gamma_{1}$, $ \gamma_{2} $ are positive and $\alpha$, $ \varkappa $  are real. Thus, the matrix $\operatorname{B}$ is self-adjoint, and  it is therefore unitarily equivalent to a real diagonal matrix
\begin{align}
	\operatorname{diag}\operatorname{B}=\operatorname{U}^{\dagger}  \,\operatorname{B}\,\operatorname{U}= \begin{bmatrix}\lambda_{1}&0\\0&\lambda_{2}\end{bmatrix}
	\nonumber
\end{align}
where for $ \mathscr{i}=1,2 $
\begin{align}
	\lambda_{\mathscr{i}}=\frac{\gamma_1+\gamma_2}{4}+(-1)^{i}\sqrt{\frac{\gamma_1^2+\gamma_2^2+2\,\kappa^2}{8}}
	\nonumber
\end{align}
Upon defining the Lindblad operators
\begin{equation}\nonumber
    \operatorname{L}_{\mathscr{j}}=\sum_{\mathscr{i}=1}^{2} \sigma_{-}^{(\mathscr{i})}\,\operatorname{U}_{\mathscr{i}\,\mathscr{j}}\,, \hspace{2.0cm}\mathscr{j}=1,2 
\end{equation}
and the self-adjoint operator
\begin{align}
	\operatorname{H}=\sum_{\mathscr{i},\mathscr{j}=1}^{2} \operatorname{A}_{\mathscr{i}\mathscr{j}}\sigma_{+}^{(\mathscr{j})}\sigma_{-}^{(\mathscr{i})}
	\nonumber
\end{align}
we finally arrive at the master equation
\begin{align}
%	\label{eq:redfield}
    \bm{\dot{\rho}}_t=-\imath\,[\operatorname{H}\,,\bm{\rho}_{t}]+\sum_{\ell=1}^{2}\lambda_{\ell}\frac{\left [\operatorname{L}_{\ell}\,,\bm{\rho}_{t}\operatorname{L}_{\ell}^{\dagger}\right ]+\left[\operatorname{L}_{\ell}\bm{\rho}_{t}\,,\operatorname{L}_{\ell}^{\dagger}\right]}{2}
    \nonumber
\end{align}

\subsection{QuTiP Implementation}

{For the numerics we assign the Lindblad weights to be
\begin{align*}
	&\Gamma_{1,t}=\gamma - 12 \exp(-2 t^3)\sin^2(15t)
\end{align*}
and
\begin{align}
	&\Gamma_{\ell,t}=\gamma\,>\,0\hspace{1.0cm}\ell=2,\dots,\mathscr{N}
	\nonumber\\
	&\Gamma_{\ell+\mathscr{N},t}=\delta\,>\,0 \hspace{1.0cm}\ell=1,\dots,\mathscr{N}
	\nonumber
\end{align}
In the time interval $ [0.2,0.25] $ we choose a solution of the over-determined system
\begin{align}
	&	\Gamma_{1,t}=\mathscr{r}_{1,t}+c_t
	\nonumber\\
	&  \delta=\mathscr{r}_{1+\mathscr{N_t}}+c_t
	\nonumber
\end{align}
by writing
\begin{align*}
	&	c_t=-b_t\,\mathscr{r}_{1+\mathscr{N},t}.
\end{align*}
with
\begin{equation}\nonumber
    b_t = \frac{(1-\textrm{sign}(\gamma_t))}{2}\frac{\gamma_t + \delta/2}{\gamma_t + \delta}
\end{equation}
The choice is merely based on the empirical observation that the resulting values of $\mathscr{r}_{1+\mathscr{N}}$,  $\mathscr{r}_{1}$ are positive and efficiently handled by QuTiP.
}

\subsection{Photo-current oscillations}
{
One way to conceptualize time-convolutionless perturbation theory is as an avenue to implement at any order in the system environment coupling constant a Markov approximation in the derivation of the master equation. The leading order corresponds to the weak coupling approximation. In that case, the Lindblad operators $ \left\{ \operatorname{L}_{\ell} \right\}_{\ell=1}^{\mathscr{L}}$ are obtained as eigenoperators 
of the unperturbed isolated system Hamiltonian $ \operatorname{H}_{\mathfrak{0}} $ \cite{BrPe2002}: 
\begin{align}
	\left[\operatorname{H}_{\mathfrak{0}}\,,\operatorname{L}_{\ell}\right]=\epsilon_{\ell}\operatorname{L}_{\ell}
	\nonumber
\end{align}
for some $ \epsilon_{l}  $'s, taking positive and negative values. We straightforwardly verify that
\begin{align}
	\left[\operatorname{H}_{\mathfrak{0}}\,,\operatorname{L}_{\ell}^{\dagger}\operatorname{L}_{\ell}\right]=0	
	\nonumber
\end{align}
immediately follows. In general, $ \operatorname{H}_{\mathfrak{0}} $
differs from the Hamilton operator $ \operatorname{H}_{t} $ in (\ref{LGKS}) by Lamb shift corrections.  
It is then reasonable to surmise that higher order corrections determined by time convolutionless perturbation theory only affect the intensity of the Lamb shift and the values of weight functions $ \Gamma_{\ell,t} $ of the Lindblad operators in (\ref{LGKS}) see e.g. \cite{SmVa2010}.  Under these hypotheses a straightforward calculation using
\begin{align}
	\mathrm{d}\operatorname{E}(\mu_{t}\nu_{\ell,t})=\Gamma_{\ell,t}\operatorname{Tr}\operatorname{L}_{\ell}\bm{\rho}_{t}\operatorname{L}_{\ell}^{\dagger} \mathrm{d}t
	\nonumber
\end{align}
allows us to derive the energy balance equation (\ref{pc:eb}).
}

{
\subsection{Measurement interpretation} 
The mathematical notion of ``instrument'' $ \mathcal{I} $
provides the description of quantum measurement adapted to quantum trajectory theory see e.g. \cite{BaBe1991}. An instrument is a map from a classical probability space $ (\Omega,\mathcal{F},\mathrm{d}P) $  \cite{Kle2005} to the space of bounded operators acting on a Hilbert space $ \mathcal{H} $ that for any pre-measurement state operator $ \bm{X} $ satisfies
\begin{subequations}
	\label{mi:instrument}
	\begin{align}
		&\label{mi:instrument1}
		\mathcal{I}_{t}(F)[\bm{X}]=\sum_{k}\int_{F}\mathrm{d}P(\omega) \mathcal{V}_{t,k}(\omega) \bm{X}\mathcal{V}_{t,k}^{\dagger}(\omega)
		\\
		&\label{mi:instrument2}
		\sum_{k}\int_{F}\mathrm{d}P(\omega) \mathcal{V}_{t,k}^{\dagger}(\omega)\mathcal{V}_{t,k}(\omega) =\operatorname{1}_{\mathcal{H} }
	\end{align}
\end{subequations}
Here $ \mathrm{d}P $ is a classical probability measure, $ F\subseteq\Omega$ is an event in the $ \sigma $-algebra $ \mathcal{F} $ describing all possible outcomes from the sample space $ \Omega $ and $\mathcal{V}_{t,k}(\omega)$ are operators acting on $ \mathcal{H} $. In (\ref{mi:instrument1}) we regard the instrument as a function of the time $ t $. 
}
{
In order to make contact with quantum trajectories unraveling a completely bounded state operator, we interpret
$ \mathcal{H} $ as an embedding Hilbert space $ \mathcal{H} = \mathcal{H}_{E} \,\otimes\,\mathcal{H}_{S} $ and $ \bm{X}=\pi(\bm{\rho}_{0}) $ as a representation of the initial state operator of the system onto $  \mathcal{H} $. Finally let $ \left\{  \operatorname{E}_{i\,j} \right\}_{i,j=1}^{\operatorname{dim}\mathcal{H}_{E}}$ be the canonical basis of the
space of operators acting on $  \mathcal{H}_{E}$. We assume $ \operatorname{dim}\mathcal{H}_{E}\,<\,\infty $ and recall that $ \operatorname{E}_{i\,j}=\bm{e}_{i}\bm{e}_{j}^{\dagger} $ where $ \left\{ \bm{e}_{i} \right\}_{i=1}^{\operatorname{dim}\mathcal{H}_{E}} $ is the canonical basis of $ \mathcal{H}_{E} $ itself.
Let now $ \operatorname{O} $ be a self-adjoint operator acting on $ \mathcal{H}_{S} $. General results
in linear operator algebra \cite{Pau2002} ensure the existence of an instrument such that for $ i\neq j $
we can write
\begin{align}
&	\operatorname{Tr}_{\mathcal{H}_{S} }\operatorname{O}\left(\bm{\Lambda}_{t\,\ti}^{(+)}(\omega)\bm{\rho}_0 \bm{\Lambda}_{t\,\ti}^{(+)\dagger}(\omega)-
	\bm{\Lambda}_{t\,\ti}^{(-)}(\omega)\bm{\rho}_0 \bm{\Lambda}_{t\,\ti}^{(-)\dagger}(\omega)\right)
	\nonumber\\
&	=\sum_{k}\operatorname{Tr}_{\mathcal{H}}\left( \frac{\operatorname{E}_{i\,j}+\operatorname{E}_{i\,j}^{\dagger}}{2}
	\,\otimes\,\operatorname{O}\,\mathcal{V}_{t\,\ti,k}(\omega) \bm{X}\mathcal{V}_{t\,\ti,k}^{\dagger}(\omega)\right)
\nonumber
\end{align}
for $  \bm{\Lambda}_{t\,\ti}^{(\pm)}$ defined in (\ref{frames}) and some $ \mathcal{V}_{t\,\ti,k}(\omega) $. We refer to \cite{Bre2004} for an explicit construction of an embedding representation. }

\section{Acknowledgments} 
We warmly thank Joachim Ankerhold, Jukka Pekola and Dmitry Golubev for useful comments and discussions. B.D. acknowledges financial support from the AtMath collaboration at the University of Helsinki. 

\section{Author Contributions}
P. M.-G. conceived the original idea. All authors equally contributed to the further refinement and development of the results.

\section{Competing Interests}
The authors declare no competing interests.

\section{Code Availability}
Code Available Upon Request.

\end{document}